\documentclass[sigconf]{acmart}
\acmConference[SE 2030]{International Workshop on Software Engineering in 2030}{November 2024}{Puerto Galinàs (Brazil)}

\usepackage{multirow,tabularx,booktabs}
\usepackage{pifont}
\usepackage{listings}
\usepackage{xcolor}
\usepackage{accsupp}

\usepackage{courier}
\usepackage{graphicx}
\usepackage{caption}
\usepackage[linesnumbered,lined,boxed,commentsnumbered]{algorithm2e}
\usepackage{caption}
\usepackage{subcaption}

\definecolor{codegreen}{rgb}{0,0.6,0}
\definecolor{codegray}{rgb}{0.5,0.5,0.5}
\definecolor{codepurple}{rgb}{0.58,0,0.82}
\definecolor{backcolour}{rgb}{0.95,0.95,0.92}
\definecolor{light-gray}{gray}{0.80}
\definecolor{celadon}{rgb}{0.67, 0.88, 0.69}
\definecolor{lawngreen}{rgb}{0.49, 0.99, 0.0}
\definecolor{lightcoral}{rgb}{0.94, 0.5, 0.5}
\colorlet{FancyVerbHighlightColor}{lime}

\lstdefinestyle{shared}{
    frame=single,
    commentstyle=\color{codegreen},
    keywordstyle=\color{magenta},
    numberstyle=\tiny\color{codegray},
    stringstyle=\color{codepurple},
    basicstyle=\ttfamily\footnotesize,
    breakatwhitespace=false,         
    breaklines=true,                 
    captionpos=b,                    
    keepspaces=true,                 
    numbers=left,                    
    numbersep=5pt,                  
    showspaces=false,                
    showstringspaces=false,
    showtabs=false,                  
    tabsize=2
}

\AtBeginDocument{%
  }

\setcopyright{acmlicensed}
\copyrightyear{2024}
\acmYear{2024}

\newcommand{\tool}{PracAPR\xspace}

\begin{document}

\title{Towards Practical and Useful Automated Program Repair \\for Debugging}

\author{Qi Xin}
\email{qxin@whu.edu.cn}
\affiliation{%
  \institution{School of Computer Science, Wuhan University\\Hubei Luojia Laboratory}
  \country{China}
}

\author{Haojun Wu}
\email{haojunwu@whu.edu.cn}
\affiliation{%
  \institution{School of Computer Science, Wuhan University}
  \country{China}
}

\author{Steven P. Reiss}
\email{spr@cs.brown.edu}
\affiliation{%
  \institution{Department of Computer Science, Brown University}
  \country{USA}
}

\author{Jifeng Xuan}
\email{jxuan@whu.edu.cn}
\affiliation{%
  \institution{School of Computer Science, Wuhan University}
  \country{China}
}
\authornote{Corresponding author}

\begin{abstract}

Current automated program repair (APR) techniques
are far from being practical and useful enough to be considered for realistic debugging.
They rely on unrealistic assumptions including the requirement of 
a comprehensive suite of test cases as the correctness criterion
and frequent program re-execution for patch validation; they are not fast; and
their ability of repairing the commonly arising complex bugs
by fixing multiple locations of the program is very limited.
We hope to substantially improve APR's practicality, effectiveness, and usefulness to help people debug.
Towards this goal, we envision \tool, 
an interactive repair system that works in an Integrated Development Environment (IDE)
to provide effective repair suggestions for debugging.
\tool does not require a test suite or program re-execution.
It assumes that the developer uses an IDE debugger
and the program has suspended at a location where a problem is observed.
It interacts with the developer to obtain a problem specification.
Based on the specification, it performs test-free, flow-analysis-based fault localization,
patch generation that combines large language model-based local repair
and tailored strategy-driven global repair, and program re-execution-free
patch validation based on simulated trace comparison to suggest repairs.
By having \tool, we hope to take a significant step towards
making APR useful and an everyday part of debugging.


\end{abstract}


\maketitle

\section{Motivation}

Programs are rarely bug-free. Debugging is an indispensable activity
in software development. It is however very costly, and can consume up to
50\% of the programming time~\cite{britton2013reversible}.
To reduce the cost of debugging and make it easier,
researchers have proposed the concept of automated program repair (APR)~\cite{monperrus2018automatic,goues2019automated,zhang2023survey,huang2023survey}
whose goal is to automatically generate a patch that
corrects a buggy program's misbehavior. 
For over a decade, more than 60 APR techniques have been developed~\cite{repair-living-review,repairtools}.
They have sought to achieve automated repair
via various strategies generally classified as
pattern-based (e.g.,~\cite{kim2013automatic,tan2016anti,liu2019tbar}),
constraint-based (e.g.,~\cite{mechtaev2016angelix,xuan2016nopol}),
search-based (e.g.,~\cite{jiang2018shaping,wen2018context}), 
and learning-based~\cite{zhang2023survey}.

Despite the promising potential, current APR techniques 
are far from being practical and useful enough to be 
integrated into an IDE for debugging.
Three key challenges remain. First, current approaches are designed
based on unrealistic assumptions. They assume the existence of a test suite
serving as the correctness criterion and require frequent program re-execution
for repair validation. 

In a realistic debugging scenario, one cannot assume the existence of a (high-quality) test suite,
especially in the initial development phase of the software.
Studies have shown that developers do not write test suites
containing a sufficient number of test cases or even do not write tests 
at all~\cite{kochhar2013empirical,beller2015and}.
As also noted by Koyuncu et al.~\cite{koyuncu2019ifixr}, bugs are often reported
without an available test suite revealing them. Surprisingly,
the bug-revealing test cases for over 90\% of the bugs in the Defects4J dataset~\cite{just2014defects4j}
were introduced after the bug was identified.

While some techniques~\cite{gao2021beyond,bader2019getafix,bavishi2019phoenix,koyuncu2019ifixr,van2018static}
have sought for test-free repair,
they are restricted to handling specific types of bugs 
(e.g., the heap-property faults~\cite{van2018static}) 
and potential issues flagged by static analyzers
and are not designed to repair general semantic bugs 
that arise while debugging.

The reliance on frequent program re-execution also makes APR not practical.
In a realistic debugging scenario,
recreating the environment for the immediate failure caused by the bug
can be difficult since the failure can be identified in a long run or in an interactive session.
Moreover, frequent program re-execution makes APR not fast.
Current approaches can take minutes (e.g.,~\cite{jiang2021cure})
or even hours to repair one bug~\cite{liu2020efficiency}.
Studies showed that developers prefer not to wait for too long~\cite{noller2022trust} for repair.
One can also imagine that APR, if integrated into an IDE 
to provide repair suggestions, is highly expected to be quick.



Second, while APR has made remarkable progress towards
local repair by generating patches addressing a single location of the program,
the repair ability is still weak.
Our statistics based on the previous evaluation of
existing tools shows that traditional non-learning-based approaches
can only repair a small fraction (less than 23\%) of
the 150 single-hunk bugs in the Defects4J v1.2 dataset.
For these bugs, the developer patches change only a single hunk of code.
Learning-based approaches, especially those using a large language model (LLM)~\cite{jiang2023impact,xia2023keep,xia2023automated,
sobania2023analysis,kang2022language,huang2023empirical},
represent a significant improvement.
However, a state-of-the-art approach Repilot~\cite{wei2023copiloting}
still failed to repair 86 (or 57.3\%) of the single-location bugs.
%
%
Moreover, existing techniques can generate spurious overfitting patches~\cite{smith2015cure,qi2015analysis,le2018overfitting,yang2022attention},
which are harmful and can adversely affect debugging~\cite{eladawy2024automated,noller2022trust}.
In summary, APR's weak ability of local repair can lower the
developer's trust of APR in dealing with even simple bugs and
make the developer unwilling to use APR for debugging.


Finally, current APR cannot do effective global repair to
tackle complex multi-location bugs whose fix requires changes for multiple locations of the program.
Zhong and Su~\cite{zhong2015empirical} found that multi-location bugs are common.
At least 40\% of real-bug fixes made by developers are used to tackle such bugs.
This finding implies that if APR is not designed to support multi-location repair,
it can have very limited usefulness.
To enhance the usability of APR, previous techniques
have attempted to address multi-location bugs using strategies
such as genetic algorithms~\cite{yuan2020toward,le2011genprog},
detection and update of evolutionary siblings~\cite{saha2019harnessing},
variational execution~\cite{wong2021varfix}, deep learning~\cite{li2022dear},
and iterative self-supervised training~\cite{ye2023iter}. 
A key problem of these approaches is that most of the repaired multi-location bugs
are actually multi-fault bugs exposed by multiple failures.
A multi-fault bug can be decomposed into multiple single-fault bugs
and is rare in realistic debugging,
as developers typically deal with one failure (fault) at a time~\cite{perez2017prevalence,ko2008debugging}.
To understand how existing approaches deal with
single-fault multi-location bugs that commonly arise while debugging,
we did a study and found that (1) about half (49.5\%) of the multi-location bugs 
in the Defects4J v1.2 dataset are multi-fault, (2)
the dataset has 118 single-fault multi-location bugs,
and (3) current techniques~\cite{yuan2020toward,saha2019harnessing,wong2021varfix,li2022dear,
ye2023iter,xia2022less,zhu2021syntax} repaired at most 8 of them.
This result shows that current APR's ability of repairing complex multi-location bugs is poor.

\section{An Envisioned Repair System for 2030}

\begin{figure*}[ht]
\centering 
\includegraphics[width=0.6\linewidth]{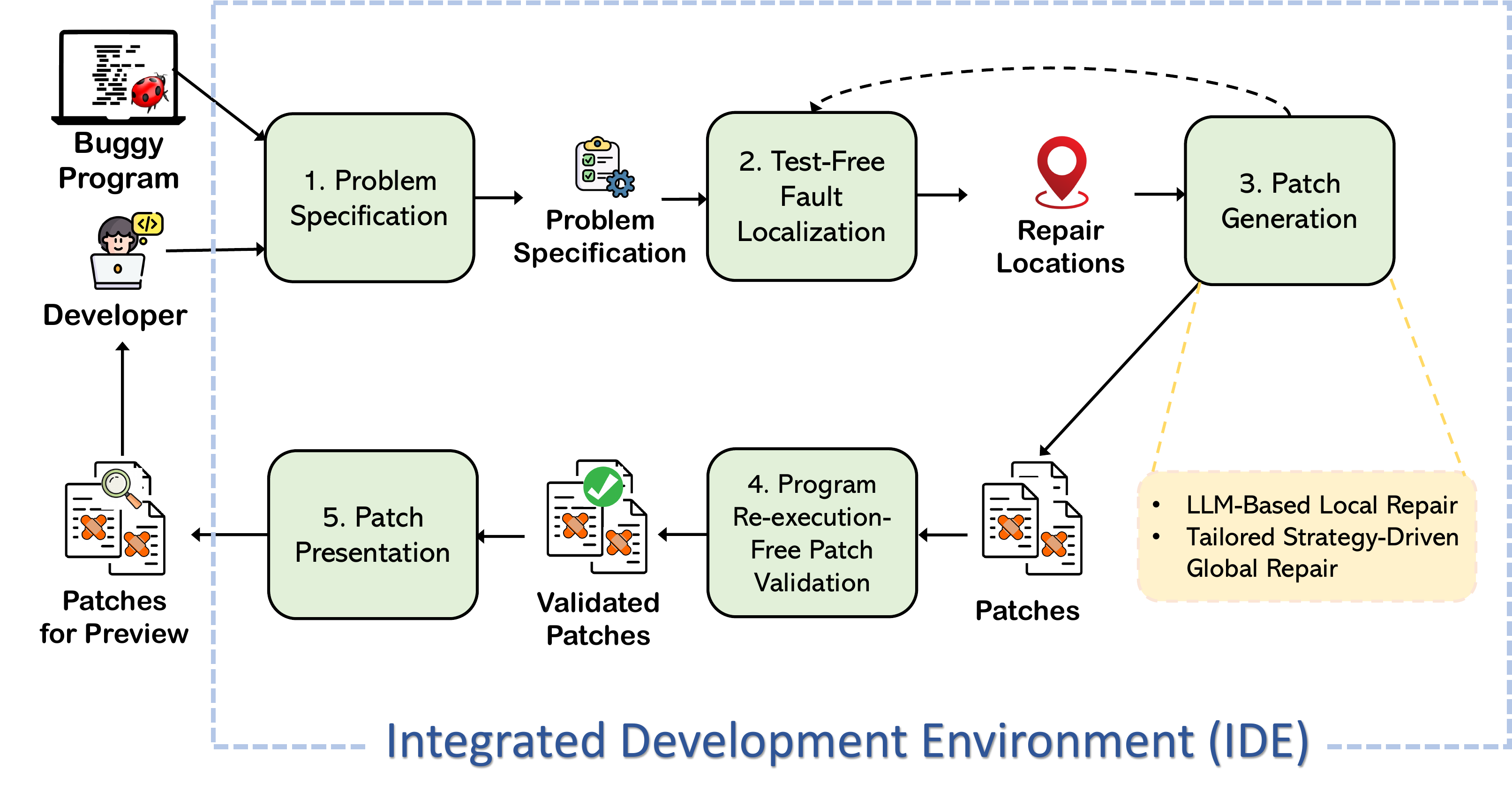}
\caption{An overview of the \tool repair system.}
\label{fig:overview}
\end{figure*}

We envision a repair system \tool that addresses the aforementioned challenges
and provides quick repair suggestions for realistic debugging.
Figure~\ref{fig:overview} shows an overview of \tool.
To overcome the unrealistic-assumption drawback,
\tool does not assume the existence of a test suite and does not require program re-execution.
It works in conjunction with an IDE debugger and
assumes that the program is stopped at a location where a problem is observed.
\tool interacts with the developer to obtain a description of the problem
and performs test-free fault localization, patch generation,
and patch validation based on the description (the problem specification) to generate repair suggestions.
Without a test suite, \tool performs flow-analysis-based
fault localization while taking into account the problem symptom,
current values from the debugger, and the current runtime stack
to compute a backward slice containing potential repair locations.
Patch generation is done with local and global repair, which we will discuss later.
Patch validation does not require program re-execution.
Instead, \tool generates via a live programming mechanism simulated traces that
reflect the real executions of the original and repaired programs 
and compares the traces to infer patch correctness. 
Finally, \tool presents the repair suggestions to the developer.
The developer can choose to preview any of the repairs
and further accept it to allow changes to be applied to the program.

For local repair, \tool uses an LLM-based approach,
aiming to improve APR's repair ability in fixing more local bugs 
(that require changes of a single location of the program)
and fixing them more precisely (by generating fewer bad patches).
To this end, \tool uses an informative prompt that 
includes the buggy location, its context, the failure input and output,
dynamic execution information (including for example the coverage and key program states),
promising patterns and fix ingredients, and user guidance
to help LLM accurately diagnose the problem and propose effective patches.
\tool also allows conversational repair and refinement and re-fixing
of the patches to improve the repair quality.

For global repair, \tool uses a variety of strategies
distilled from our analysis of multi-location patches
that led to a characterization of 8 types of partial patch relationships.
The strategies include performing iterative repair to generate patches
location by location to address bugs that require fixing different issues,
performing simultaneous repair to address related issues,
using local repair to tackle single-location-alike bugs,
and using pattern-based methods to generate other common patches 
that involve for example adding a definition and use of
a variable and inserting if-wrap code (i.e., an if-statement wrapping a code hunk).

\section{Ongoing and Future Work}

We next discuss our ongoing and future work for realizing \tool.

\subsection{Interactive Test-Free Repair Framework}

We have developed an interactive test-free repair framework ROSE~\cite{reiss2023quick},
which provides initial solutions for fault localization and patch validation 
without requiring a test suite and program re-execution.

ROSE works in the Eclipse-based Code Bubbles IDE~\cite{bragdon2010code}
and allows easy integration of existing APR patch generators.
ROSE assumes that the program is suspended at a location where unexpected behavior is observed.
It interacts with the developer to obtain a problem description.
The developer can specify that an exception is unexpected,
a line should not be executed, or a variable should not hold a certain value.
Based on the specification, ROSE performs test-free fault localization
using an abstract-interpretation-based flow analysis to statically compute 
a partial backward slice containing potential repair locations.
It invokes the patch generators that have been plugged into the framework
to make patches for those locations. For patch validation without using test cases 
or allowing dynamic program re-execution, ROSE generates simulated traces 
based on a live-programming system SEEDE~\cite{reiss2018seede} for both the original and repaired executions 
and then compares the traces with respect to the problem to infer patch correctness.
Finally, ROSE presents a limited number of prioritized patches.
The developer can choose a patch for a preview, which highlights the code
before and after the repair with differences, and ask ROSE to make the repair.
More details can be found in~\cite{reiss2023quick,reiss2022quick}.


We evaluated the effectiveness and utility of ROSE with a repair experiment and a user study. 
Our results showed that ROSE's test-free fault localization and patch validation 
are highly effective: the fault localization included the correct repair location
for 89\% of the bugs tested and the patch validation gave a top-5 rank for 
all correct repairs; that a ROSE-based tool can repair as many as 36/40 QuixBugs and 
37/60 Defects4J bugs in only seconds; and that ROSE helped 44\% more participants
succeed in a debugging task and helped reduce the debugging time by about 16.5\%.
Overall, we believe that ROSE is a promising repair framework
that can make debugging easier.

We plan to build \tool on top of ROSE, and we see two ways for improvement.
First, we want to improve ROSE's user interaction for problem specification 
by exploring not only a better presentation of the failure information to 
facilitate understanding of the program semantics and the failure but also 
more forms of the specification (based on for example constraints 
and even natural language) to effectively guide fault localization and patch validation.
Second, we want to investigate learning-based trace comparison 
while considering more information about the execution to enhance patch validation. 
\subsection{LLM-Based Local Repair}
\label{sec:local_repair}

The LLM has demonstrated superior abilities in repairing software bugs~\cite{xia2023keep,xia2023automated}.
We believe that an LLM-based approach is promising in generating high-quality local patches
(addressing single locations of the program for repair).
Since ChatGPT is widely recognized as a prominent LLM for tackling various software engineering tasks,
we consider a ChatGPT-based approach
that serves as a key component of \tool
to accurately infer the problem and provide a low number of promising patches
for effective local repair.


We are conducting a study to understand the failure of ChatGPT-based
approaches~\cite{xia2023keep,xia2023automated} and motivate possible ways for improvement.
We seek to answer three research questions: 
(1) What are the characteristics of the bugs that ChatGPT fails to repair?
(2) What are the most common mistakes that ChatGPT makes?
and (3) How to improve ChatGPT to repair more bugs?

Our current result shows that existing approaches
are weak in that they use prompts that include only the buggy location,
its limited context, and shallow information about the failure
(including the input and the failing assertion). 
This is often insufficient for ChatGPT to understand the program semantics
and the problem and can result in incorrect patches raising new problems.


\begin{figure}

\begin{lstlisting}[
  frame=lines,
  breaklines=true,
  basicstyle=\ttfamily\scriptsize,
  xleftmargin=0em,
  numbers=left,
  numbersep=0.1em,
  language=java,
  escapeinside={\%*}{*)}
]
public TimeSeries createCopy(int start, int end)
            throws CloneNotSupportedException {
    if (start < 0) {
        throw new IllegalArgumentException("Requires start >= 0.");
    }
    if (end < start) {
        throw new IllegalArgumentException("Requires start <= end.");
    }
    TimeSeries copy = (TimeSeries) super.clone();
    %*\colorbox{green!20}{+ copy.minY = Double.NaN;}*)
    %*\colorbox{green!20}{+ copy.maxY = Double.NaN;}*)
    copy.data = new java.util.ArrayList();
    if (this.data.size() > 0) {
        for (int index = start; index <= end; index++) {
            TimeSeriesDataItem item
                    = (TimeSeriesDataItem) this.data.get(index);
            TimeSeriesDataItem clone = (TimeSeriesDataItem) item.clone();
            try {
                copy.add(clone);
            }
            catch (SeriesException e) {
                e.printStackTrace();
            }
        }
    }
    return copy;
}
\end{lstlisting}
\caption{Patch for the Chart\_3 bug.}
\label{fig:chart3patch}
\end{figure}

Figure~\ref{fig:chart3patch} shows for example the buggy method for
the Defects4J Chart\_3 bug and the patch (lines 10 and 11).
To repair the bug, a state-of-the-art ChatGPT-based approach 
ChatRepair~\cite{xia2023keep} uses a prompt that includes the code of the buggy method, 
the name of the failing test case \texttt{testCreateCopy3},
the failing assertion \texttt{assertEquals(101.0, s2.getMaxY(), EPSILON)},
and the failure message \texttt{expected:<101.0> but was:<102.0>}.
It does not however inform ChatGPT of the test input triggering the failure.
Nor does it describe the behavior of the invoked method \texttt{add} (line 19)
showing how \texttt{minY} and \texttt{maxY} are updated 
(key information for bug understanding) and provide the details about the failure execution.
Due to insufficient knowledge of the failure, ChatGPT's problem diagnosis is 
shallow and inaccurate -- it thought that there is a problem
with the loop copying the data and did not seem to understand 
that it was the update of the \texttt{minY} and \texttt{maxY} values
that causes the error. As a result, ChatGPT proposed a patch
changing the loop condition (line 14), which is incorrect.

To help ChatGPT understand the failure,
we plan to use an augmented prompt that includes not only what ChatRepair uses in its prompt
but also the code of the failing test case (including the test input), 
the definition of related methods (including \texttt{add}), and the execution trace 
showing not only what lines are exercised in the failure run and their order 
but also the key program state (variable and field values). 
Using a prompt like this, ChatGPT successfully understands
that the failure is related to ``how the min and max y values are updated
after copying a subset''. This finally leads to a patch that correctly
updates the min and max y values.

An augmented prompt, even with more failure and execution information,
may not necessarily help ChatGPT figure out what is wrong. 
And even if ChatGPT precisely understands the problem, it may still 
fail to generate the correct patch tackling the problem in the right way.
For example, ChatGPT may know that there is an invalid case
where the \texttt{start} index is greater than \texttt{end}
but can be unsure about how to process it -- whether the program should throw an exception, 
return a special value, or do something else.
One way to mitigate this problem is to solicit user feedback
showing for example an exception is expected, a certain line should not be executed,
or a variable should not hold a value.

In addition to using augmented prompts, we will also
explore combining ChatGPT with traditional pattern-based and search-based methods 
(finding for example effective patterns and fix ingredients) to guide the repair,
performing conversational repair highlighting the (negative) influence of
the previous patches to allow ChatGPT to reflect on its mistakes
for improvement, and conducting post-processing operations 
refining and re-fixing the patches to improve repair quality.


\subsection{Global Repair Driven by Tailored Strategies}

Existing global repair techniques have adopted various strategies
for multi-location repair. 
The evaluation of these techniques is however based on the Defects4J bug dataset~\cite{just2014defects4j}
and is severely misguided, as the dataset is filled with multi-fault bugs.
Multi-fault bugs can be decomposed into independent single-fault bugs
triggering different failures. Repairing multi-fault bugs by handling
multiple failures simultaneously is practically uncommon for debugging,
as the developer typically deals with one failure at a time~\cite{perez2017prevalence,ko2008debugging}.

To understand existing approaches' abilities of
repairing single-fault multi-location bugs, we proposed an approach 
to detect such bugs and found that there are 118 single-fault multi-location bugs
in the Defects4J v1.2 dataset and that current approaches~\cite{yuan2020toward,saha2019harnessing,wong2021varfix,li2022dear,
ye2023iter,zhu2021syntax,xia2022less} repaired at most 8 bugs, suggesting weak repair abilities.

We aim to design a global repair approach that can effectively address single-fault multi-location bugs.
Towards this goal, we went about analyzing the developer (ground-truth) patches
for a sample of the single-fault multi-location bugs (about one third, or 75 in total).
We wanted to understand why the repair needs to addresses multiple locations,
what are the characteristics of the partial patches made at different locations,
and furthermore what strategies to consider for patch generation
based on the characteristics.

The analysis has led to a characterization of 8 partial patch relationships
summarized below.
\vspace{-2pt}
\begin{itemize}
\item \textbf{DU}: Partial patches with this relationship add the definition of 
variables, fields, packages, or methods and later use what has been defined for repair.
\item \textbf{OA}: Partial patches with this relationship can be done in one 
repair action or operation by for example adding an if-statement wrapping a code hunk.
\item \textbf{RIF}: This relationship indicates that the partial patches 
are used to address related issues that arise in different locations.
\item \textbf{DIF}: This relationship indicates that the partial patches 
address different issues arising from different program parts that may implement the same functionality.
\item \textbf{EOH}: Partial patches with this relationship can be considered 
as a single-hunk patch for reasons such as that there is only one partial patch 
that is semantically needed and the others are created only to improve readability.
\item \textbf{SU}: In this relationship, some partial patches are created to do the setup work
by updating a variable, field, or method while the others use what has been updated for repair.
\item \textbf{ONPF}: In this relationship, some partial patches can fix the original problem 
and resolve the original failure. Unfortunately, they also raise new problems 
triggering new failures, which can be tackled by other partial patches.
\item \textbf{FU}: Some partial patches serve as the primary changes for
correcting the misbehavior of the program. Others are needed to undo the negative influence 
brought by the previous changes.
\end{itemize}
\vspace{-2pt}

As the next step, we plan to design specialized repair strategies
based on the relationships and develop a global repair approach 
that uses these strategies to obtain guided exploration for effective multi-location repair. 
An approach that we envision to have uses the LLM-based method
discussed in Section~\ref{sec:local_repair} to generate single-location patches.
It performs iterative repair via repeated single-location-based fault localization 
and patch generation to generate patches of the DIF, ONPF, and FU relationships.
Unlike existing approaches~\cite{yuan2020toward,ye2023iter},
our approach considers a variety of program syntactic and semantic features
and execution information to infer promising patches for further evolution.
To generate the RIF patch, our approach reuses a simultaneous strategy~\cite{saha2019harnessing}
that identifies locations for co-evolution and applies similar changes 
to those locations. The approach relies on local repair to address EOH
and uses pattern-based methods to generate DU, SU, and OA patches.

We envision to have a suite of specialized patch generators.
Once a failure occurs, one would not easily know which generators to use
for repair but can run all the generators in parallel to
get all the patches. This can be further improved via a trained multi-classifier~\cite{meng2022improving,aleti2021apr}
to select the most suitable generators or an ensemble approach
(e.g.,~\cite{zhong2023practical}) for generator prioritization.

\bibliographystyle{ACM-Reference-Format}
\bibliography{paper}
\balance

\end{document}